\documentstyle[preprint,aps]{revtex}

\newcommand{\be}{\begin{equation}}
\newcommand{\ee}{\end{equation}}
\newcommand{\bea}{\begin{eqnarray}}
\newcommand{\eea}{\end{eqnarray}}

\input psfig

\tightenlines

\begin{document}

\draft
%@twocolumnfalse\endcsname

\title{\bf DIMER ORDER WITH STRIPED CORRELATIONS\\
IN THE $J_1$-$J_2$ HEISENBERG MODEL}

\author{Rajiv R. P. Singh$^1$\cite{byline1}, Zheng Weihong$^2$\cite{byline2},  
C.J. Hamer$^2$\cite{byline3}, and J. Oitmaa$^2$\cite{byline4}}
\address{
$^1$Department of Physics, University of California, Davis, CA 95616; \\
$^2$School of Physics,
The University of New South Wales, Sydney, NSW 2052, Australia.
}

%\date{May 22, 1995}
\date{\today}

\maketitle

\begin{abstract}
Ground state energies for plaquette and dimer order in the $J_1\mbox{-}J_2$ square-lattice
spin-half Heisenberg model are compared using series expansion methods.
We find that these energies are remarkably close to each other at intermediate
values of $J_2/J_1$, where the model is believed to have a quantum disordered
ground state. 
They join smoothly with those obtained from the Ising expansions 
for the 2-sublattice N\'eel-state at $J_2/J_1 \approx 0.4$,
suggesting a second order transition from a N\'eel state to a quantum disordered state,
whereas they cross the energy for the
4-sublattice ordered state at $J_2/J_1 \approx 0.6$ at a large angle,
implying a first order transition to
the 4-sublattice magnetic state.
The strongest evidence that the plaquette phase is {\it not} realized in
this model comes from the analysis of the series for the singlet and triplet excitation
spectra, which suggest an instability in the plaquette phase. Thus, our study
supports the recent work of Kotov et al, which presents a strong picture
for columnar dimer order in this model. We also discuss the striped nature
of spin correlations in this phase, with substantial resonance all along columns of dimers.
\end{abstract}

%\phantom{.}
%]

\narrowtext

\section{Introduction}
There has been considerable study, over the last decade, of the
frustrated spin-$\case 1/2$ square lattice Heisenberg antiferromagnet
(the ``$J_1\mbox{-}J_2$ antiferromagnet"). These studies include exact diagonalizations
on small systems\cite{dag89,sin90,sch92,ric93},
spin-wave calculations\cite{cec92,dot94}, series expansions\cite{gel89,Isingexp},
and a field-theoretic large-$N$ expansions\cite{read91}.

These studies, and others, have provided a substantial body of evidence that
the ground state of this system, in the region $0.4 \lesssim J_2/J_1 \lesssim 0.6$,
has no long-range magnetic order and has a gap to spin excitations.
For $J_2/J_1 \lesssim 0.4$ the model has conventional 
antiferromagnetic N\'eel order whereas for $J_2/J_1 \gtrsim 0.6$
the system orders in a columnar $(\pi,0)$ phase. Whether this
``intermediate phase" is a spatially homogeneous spin-liquid,
or whether it has some type of spontaneously broken symmetry leading to
a more subtle type of long-range order, has not been conclusively
established.

Zhitomirsky and Ueda\cite{zhi96} have proposed a plaquette resonating valence bond (RVB) 
phase, which breaks translational symmetry along both $x$ and $y$ axes,
but preserves the symmetry of interchange of the two axes.
The horizontal and vertical dimers resonate within a plaquette. An early
series study\cite{gel90} had investigated the relative stability of 
various spontaneously dimerized states and had concluded that a columnar
dimerized phase was the most promising candidate for the intermediate
region, in agreement with the large-$N$ expansions.
Zhitomirsky and Ueda\cite{zhi96} claim their plaquette phase has a lower
energy than this columnar dimer phase, but we find this to be incorrect.

Further support for the columnar dimer scenario comes from recent work
of Kotov {\it et al.}\cite{dimerexp}, who combine an analytic many body theory with 
extended series and diagonalization results to study the nature and
stability of the excitations in the intermediate region. It is
argued that where the N\'eel phase becomes unstable the system
will develop not only a gap for triplet excitations but also a gapped
low-energy singlet which reflects the spontaneous symmetry breaking.
This is clearly seen in the calculations. At $J_2/J_1 \simeq 0.38$ a 
second order transition occurs, with the energies of N\'eel phase and dimerized phase joining
smoothly, and the energy gap and dimerization vanishing.

It is the aim of this paper to further investigate, using series methods,
the competing possibilities of columnar dimerization versus plaquette order
in the intermediate region of the $J_1\mbox{-}J_2$ antiferromagnet. It is 
conceivable that both occur, with a transition from one
 to the other. However, such a transition, reflecting a change
of symmetry, is expected to be first-order and not well suited to series methods.
If both phases are locally stable the most direct way to compare them is 
by comparison of the ground state energies. If one is unstable this should
show up by the closing of an appropriate gap or by the divergence of an
appropriate susceptibility. In this paper we calculate the ground
state energy and the singlet and triplet excitation spectra by series expansions
about a disconnected plaquette Hamiltonian. We also calculate the
susceptibility for the dimer phase to break translational symmetry in the direction 
perpendicular to the dimers. This susceptibility will be large if there is substantial
resonance in the dimer phase and will diverge if there is an 
instability to the plaquette RVB phase.

Combining the plaquette expansion results with the dimer expansions of Kotov 
{\it et al.}\cite{dimerexp}, a very interesting picture emerges for
the quantum disordered phase. We find that the plaquette phase is 
unstable and hence is not the ground state for this model. The
dimer phase, on the other hand, is stable. However, there is substantial
resonance in the dimer phase. The spin-spin correlations are
not simply those of isolated dimers. Instead, the nearest neighbor correlations are nearly
identical along the rungs and chains of dimer columns. In contrast, the
correlations from one dimer column to the next are much weaker. The spin-gap phase
appears separated from the N\'eel phase by a second order transition, whereas it is
separated from the columnar phase by a first order transition. These results
are in remarkable agreement with the large-N theories \cite{read91}. The existence
of a quantum critical point separating an antiferromagnetic phase and
a quantum disordered phase with striped correlations in a microscopic model
makes this critical point a particularly interesting one. The role of doping
and its implications for high-$T_c$ materials deserves
further attention.

\section{Series Expansions and Results}
We study the Hamiltonian 
\begin{equation}
H = J_1 \sum_{{\rm n.n.}} {\bf S}_i\cdot {\bf S}_j  +
 J_2 \sum_{{\rm n.n.n.}} {\bf S}_i\cdot {\bf S}_j \label{H}
\end{equation}
where the first sum runs over the nearest neighbor and the second over the
second nearest neighbor spin pairs of the square-lattice.
We denote the ratio of couplings as $y= J_2/J_1$.
The linked-cluster expansion method has been
previously reviewed in several articles\cite{he90,gel,gelmk},
and will not be repeated here. 
To carry out the series expansion about 
the disconnected-plaquette state for this system, 
we take the interactions denoted by the
thick solid and dashed bonds in Fig. 1 as the unperturbed Hamiltonian,
and the rest of the interactions as a perturbation. That is, we define
the  following Hamiltonian 
\begin{equation}
H = H_0 + H_1
\end{equation}
where the unperturbed Hamiltonian ($H_0$) and perturbation ($H_1$) are
\begin{eqnarray}
H_0 &=& J_1 \sum_{\langle ij\rangle \in A} {\bf S}_i \cdot {\bf
S}_j +
 J_2 \sum_{\langle ij\rangle \in B} {\bf S}_i\cdot {\bf S}_j \nonumber \\ & & \\
H_1 &=& \lambda J_1 \sum_{\langle ij\rangle \in C} {\bf S}_i\cdot {\bf
S}_j  +
 \lambda J_2 \sum_{\langle ij\rangle \in D} {\bf S}_i\cdot {\bf S}_j \nonumber
\end{eqnarray}
and the summations are over intra-plaquette nearest-neighbor bonds
(A), intra-plaquette second nearest-neighbor bonds (B),
inter-plaquette nearest-neighbor bonds (C), inter-plaquette second
nearest-neighbor bonds (D), shown in Fig. 1. With this Hamiltonian, one 
can carry out an expansion in powers of $\lambda$, and 
at $\lambda=1$ one recovers the
original Hamiltonian  in Eq. (\ref{H}). Thus, although we expand about
a particular state, i.e. a plaquette state, our results at $\lambda=1$ describe
the original system without broken symmetries, provided no intervening
singularity is present. Such perturbation expansions about an unperturbed
plaquette Hamiltonian have been used previously to study Heisenberg models
for CaV$_4$O$_9$\cite{cavoser}.

It is instructive to consider the states of an isolated plaquette.
There are two singlet states, one with energy $(-2+y/2) J_1$ and
the other with energy $(-3y/2) J_1$. The former is the ground state
for $y<1$ and corresponds to pair singlets resonating between the vertical
and horizontal bonds of the plaquette. It is even under a $\pi/2$ rotation.
The latter is the ground state for $y>1$ and is odd under a $\pi/2$ rotation.
The wavefunctions for these two singlet states are
\bea
\psi_1 &=& {1\over \sqrt{12}} {\Big [}
 {\Big (} \begin{array}{cc} + & + \\ - & - \end{array} {\Big )}
+ {\Big (} \begin{array}{cc} + & - \\ + & - \end{array} {\Big )}
+ {\Big (} \begin{array}{cc} - & - \\ + & + \end{array} {\Big )}
+ {\Big (} \begin{array}{cc} - & + \\ - & + \end{array} {\Big )}
-2 {\Big (} \begin{array}{cc} + & - \\ - & + \end{array} {\Big )}
-2 {\Big (} \begin{array}{cc} - & + \\ + & - \end{array} {\Big )}
{\Big ]} \nonumber \\
&=& {1\over \sqrt{3}} {\Big [}
 {\Big (} \begin{array}{c} 
\begin{picture}(30,10)(-5,-5) 
\put(0,0) {\circle{6}}
\put(3,0){\line(1,0){12}}
\put(18,0){\circle{6}}
\end{picture}
\\
\begin{picture}(30,10)(-5,-5) 
\put(0,0) {\circle{6}}
\put(3,0){\line(1,0){12}}
\put(18,0){\circle{6}}
\end{picture}
\end{array} {\Big )}
+
{\Big (} \begin{array}{cc} 
\begin{picture}(10,30)(-5,-5) 
\put(0,0) {\circle{6}}
\put(0,3){\line(0,1){12}}
\put(0,18){\circle{6}}
\end{picture}
&
\begin{picture}(10,30)(-5,-5) 
\put(0,0) {\circle{6}}
\put(0,3){\line(0,1){12}}
\put(0,18){\circle{6}}
\end{picture}
\end{array} {\Big )}
{\Big ]} \\
\psi_2 &=& {1\over 2} {\Big [}
 {\Big (} \begin{array}{cc} + & + \\ - & - \end{array} {\Big )}
- {\Big (} \begin{array}{cc} + & - \\ + & - \end{array} {\Big )}
+ {\Big (} \begin{array}{cc} - & - \\ + & + \end{array} {\Big )}
- {\Big (} \begin{array}{cc} - & + \\ - & + \end{array} {\Big )}
{\Big ]} \nonumber \\
&=&  {\Big [}
{\Big (} \begin{array}{cc} 
\begin{picture}(10,30)(-5,-5) 
\put(0,0) {\circle{6}}
\put(0,3){\line(0,1){12}}
\put(0,18){\circle{6}}
\end{picture}
&
\begin{picture}(10,30)(-5,-5) 
\put(0,0) {\circle{6}}
\put(0,3){\line(0,1){12}}
\put(0,18){\circle{6}}
\end{picture}
\end{array} {\Big )}
-
 {\Big (} \begin{array}{c} 
\begin{picture}(30,10)(-5,-5) 
\put(0,0) {\circle{6}}
\put(3,0){\line(1,0){12}}
\put(18,0){\circle{6}}
\end{picture}
\\
\begin{picture}(30,10)(-5,-5) 
\put(0,0) {\circle{6}}
\put(3,0){\line(1,0){12}}
\put(18,0){\circle{6}}
\end{picture}
\end{array} {\Big )}
{\Big ]} 
\eea
where \begin{picture}(25,6)(-3,-3) 
\put(0,0) {\circle{6}}
\put(3,0){\line(1,0){12}}
\put(18,0){\circle{6}}
\end{picture}
means these two spins form a singlet.
There are three triplet states, one with energy $(-1+y/2) J_1$ and
a degenerate pair with energy $(-y/2) J_1$; like the singlets, these
have a level crossing at $y=1$. Under a $\pi/2$ rotation
the former is odd, while the latter two are even  and odd, respectively.
% while the latter transform into each other (or they can be
% transformed to have even  and odd under $\pi/2$ rotation).
Finally there is a quintuplet state at $(1+y/2) J_1$, which is even
under a $\pi/2$ rotation. For $y<1/2$ and $y>2$ the first excited state of
the plaquette is a triplet, while for $1/2<y<2$ it is the other singlet.
These states and corresponding energies are shown in Figure 2.
The eigenstates of $H_0$, the unperturbed Hamiltonian, are direct
products of these plaquette states.

To derive the plaquette expansions we identify each plaquette as a 
16 state quantum object, and these lie at the sites of a square lattice
with spacing $2a$, where $a$ is the original lattice spacing. Interactions
between plaquettes connect first and second-neighbor sites on this new
lattice. The cluster data is thus identical to that used by us previously\cite{Isingexp}
to derive Ising expansions for this model. Because there are 16 states at each cluster
site, the vector space grows very rapidly with the number of sites and thus 
limits the maximum attainable order for plaquette expansions to
considerably less than can be achieved for dimer or Ising expansions.

We have computed the ground state energy $E_0$ to order $\lambda^7$, for fixed values of
the coupling ratio $y$. The series are analysed using integrated differential
approximants\cite{gut}, evaluated at $\lambda=1$ to give the ground
state energy of the original Hamiltonian. The estimates with error bars representing
confidence limits, are shown in Figure 3. For comparison we also show previous results
obtained from Ising expansions\cite{Isingexp} and dimer expansions\cite{dimerexp}.
We find that, in the intermediate region, the ground state energy for both plaquette and
dimer phase are very close to each other and cannot be used to distinguish between
them. The dimer expansion yields slightly lower energies near the transition to the N\'eel
phase. We do not draw any conclusions from this.

Zhitomirsky and Ueda\cite{zhi96} have claimed that the ground state energy from a 
second-order plaquette expansion is -0.63 (at $y=0.5$), much lower that the dimer expansion result 
-0.492.  This result appears incorrect. At $J_2/J_1=\case 1/2$ the ground
state energy is given by
\bea
4 E_0/NJ_1 &=& -7/4 - 277 \lambda^2/1440 -0.001357 \lambda^3 -0.0210609 \lambda^4 \nonumber \\
&& -0.000319586 \lambda^5 -0.00580643 \lambda^6 -0.001822686 \lambda^7 +O(\lambda^8)
\eea
The second order result (at $\lambda=1$) is $E_0/NJ_1 = -0.485$,
rather than $-0.63$. We note that if the second order coefficient were
4 times larger then the resulting energy would be $-0.62986$.

We have also derived series, to order $\lambda^6$, for the
singlet and triplet excitation energies,
$\Delta_s(k_x,k_y)$, $\Delta_t(k_x,k_y)$ using the method
of Gelfand\cite{gel}, and taking as unperturbed eigenfunctions
the corresponding plaquette states. The low order terms for
$J_2/J_1=0.5$ are given by:
\bea
\Delta_s(k_x,k_y)/J_1 &=& 1 -301\lambda^2/1440  + 137 \lambda^3/86400 
+ 217 \lambda^3 \cos(k_x) \cos(k_y) /172800\nonumber \\
&& +(-5\lambda^2/16 - 89 \lambda^3/9600 ) [\cos(k_x)+\cos(k_y)]/2 
\\
\Delta_t(k_x,k_y)/J_1 &=& 1 - 3691 \lambda^2/30240   
+ (- 2 \lambda/3  + 11 \lambda^2/720 )  [\cos(k_x)+\cos(k_y)]/2 \nonumber \\
&& -\lambda^2  [\cos(2 k_x)+\cos(2 k_y)]/120 
+ (\lambda/3   -5\lambda^2/96   ) \cos(k_x) \cos(k_y) \nonumber \\
&& - \lambda^2  [\cos( 2 k_x) \cos( k_y) + \cos(k_x) \cos( 2k_y)]/90
+ 7 \lambda^2 \cos( 2 k_x) \cos( 2 k_y)/360
\eea
The full series are available on request.
We first consider the triplet excitations. Figure 4 shows $\Delta_t (k_x,k_y)$ along
high symmetry directions in the Brillouin zone for $\lambda=0.5$ and various coupling
ratios $y$. For $\lambda \lesssim 0.6$ the series are well converged and direct summation and
integrated differential approximants give essentially
identical results. We find that the minimum gap occurs at $(0,0)$
for $J_2/J_1 \lesssim 0.55$ and  moves to $(\pi,0)$ for $J_2/J_1\gtrsim 0.55$.
Next we seek to locate the critical point $\lambda_c$ where the triplet gap vanishes.
This is done using Dlog Pad\'e approximants to the gap series at the 
appropriate $(k_x,k_y)$. In practice this works well when the minimum gap
lies at $(0,0)$. For $J_2=0$ we find a critical point at
$\lambda_c=0.555(10)$. We can compare this result with recent work of Koga 
{\it et al.}\cite{koga98} who obtain $\lambda_c = 0.112$ from a modified spin-wave
theory and $\lambda_c\simeq 0.54$ from a 4th order plaquette expansion.
The critical point $\lambda_c$ increases with increasing $y$. At $y=0.5$,
at the approximate centre of the intermediate phase, we find $\lambda_c \simeq 0.89(7)$.
This result has some uncertainty but, if accurate, means that the plaquette
phase becomes unstable before the full
Hamiltonian ($\lambda=1$) is reached. The associated critical exponent $\nu$ describing
the vanishing of the triplet gap is about 0.7 for $J_2/J_1 < 0.4$, 
suggesting that the transition
lies in the universality class of the classical $d=3$ Heisenberg model. On
the other hand, for $J_2/J_1 \gtrsim 0.4$ the exponent $\nu$ is about 0.4.
This supports the existence of an intermediate phase lying in a different
universality class.

Figure 5 shows the singlet excitation energy $\Delta_s (k_x,k_y)$ along  
high symmetry directions in the Brillouin zone for $\lambda=0.5$ and the same coupling
ratios $y$ as Figure 4. Again the series are well converged and direct summation and
integrated differential approximants give essentially
identical results. We find that the minimum gap occurs at $(0,0)$
for all $J_2/J_1$. We have also noted that for $J_2/J_1=0.5$, the triplet excitation and
the singlet excitation have same gap at $\lambda=0$, but at $\lambda=0.5$, the singlet
gap is considerable larger than the triplet gap, this means probably that the triplet gap 
close before the singlet gap at $J_2/J_1=0.5$. The critical point obtained 
by the Dlog Pad\'e approximant to the singlet gap is also
generally slightly larger than that obtained from the triplet gap
around $J_2/J_1=0.5$ (see Fig. 6). 

The full phase diagram in the parameter space of $J_2/J_1$ and $\lambda$ could be
very interesting from the point of view of quantum phase transitions,
but may not be easy to determine by numerical methods. 
Some possible scenarios are shown in Fig. 7.
One possibility is that the plaquette  phase, for all $J_2/J_1$, has  an instability
to some magnetic phase and the dimerized phase exists only very close to
$\lambda=1$ inside the magnetic phases. A second possibility is that the
plaquette-N\'eel critical line meets the N\'eel-dimer critical line at some
multicritical point at a value of $J_2/J_1$ around $0.5$, after which
there is a first order transition between the plaquette and the dimer
phases. A third possibility is that the plaquette-N\'eel, N\'eel-dimer
and plaquette-columnar critical lines all meet at some multicritical
point. The numerically determined phase diagram is particularly
uncertain in the interesting region, $0.5\lesssim J_2/J_1\lesssim 0.6$.,
where incommensurate correlations could also become important.

Lastly we have derived expansions for a number of generalized susceptibilities.
These are defined by adding an appropriate field term
\be
\Delta H = h \sum_{ij} Q_{ij}
\ee
to the Hamiltonian
and computing the susceptibility from
\be
\chi_Q = - {1\over N} \lim_{h\to 0} {\partial^2 E_0(h) \over \partial h^2}
\ee
A divergence of any susceptibility signals an instability
of that phase with respect to the particular type of order 
incorporated in $\chi$.

We have computed two different susceptibilities from the plaquette expansion. One is the
antiferromagnetic (N\'eel) susceptibility with the  operator $Q_{ij}$
\be
Q_{i,j} = (-1)^{i+j} S^z_{i,j}
\ee
The other is the dimerization susceptibility with the operator $Q_{ij}$
\be
Q_{i,j} = {\bf S}_{i,j}\cdot {\bf S}_{i+1,j} - {\bf S}_{i,j}\cdot {\bf S}_{i,j+1}
\ee
which breaks the symmetry  of interchange of  $x$ and $y$ axes.
We have computed series to order $\lambda^5$ for the antiferromagnetic  susceptibility 
and  to order $\lambda^4$ for the dimerization
susceptibility.
% From the dimer expansion we have computed the N\'eel, columnar and 
% plaquette susceptibilities to order $\lambda^7$. These are listed in Table I, together
% with the appropriate operator $Q_{ij}$.
The series  have been analyzed by Dlog Pad\'e approximants.
The series for the antiferromagnetic susceptibility 
shows the same critical points (within error bars) as those obtained from the triplet gap for 
$J_2/J_1 \lesssim 0.4$. The series for the dimerization susceptibility 
is very irregular, and does not yield useful results.
For example, for $J_2/J_1=0.5$, the series is:
\be
\chi_d = 629/90 + 101\lambda /300  + 2.0097647 \lambda^2
 -0.269629 \lambda^3 + 0.438527 \lambda^4 +O(\lambda^5)
\ee

For completeness, we also compute the susceptibility for the dimer phase to become unstable
to the plaquette phase from an  expansion about
isolated columnar dimers\cite{dimerexp}, by adding the following
field term:
\be
\Delta H = h \sum_{i,j} (-1)^j {\bf S}_{i,j}\cdot {\bf S}_{i,j+1}
\ee
which breaks the translational symmetry in the direction perpendicular to
the dimers.
The series has been computed up to order $\lambda^7$, (note that $\lambda$ here
is the parameter of dimerization). 

An analysis of the series shows that this
susceptibility becomes very large as $\lambda\to 1$, for all $J_2/J_1$
and the critical $\lambda$,
where the susceptibility appears to diverge, approaches unity from above
as $J_2/J_1$ is increased to $0.5$. This implies that there are
staggered bond correlations in the direction perpendicular
to the dimers, which extend over a substantial range. An interesting question is,
in the absence of the plaquette phase as discussed earlier, what could these
correlations represent? At this stage it is useful to recall another calculation 
by Kotov {\it et al.}\cite{dimerexp}. Within the dimer expansion, they
calculated two different dimer order parameters,
\be
D_x = |<{\bf S}_{i,j}\cdot {\bf S}_{i+1,j}>-
<{\bf S}_{i+1,j}\cdot {\bf S}_{i+2,j}>|,
\ee
and,
\be
D_y = |<{\bf S}_{i,j}\cdot {\bf S}_{i+1,j}>-
<{\bf S}_{i,j}\cdot {\bf S}_{i,j+1}>|,
\ee
where the elementary dimers connect spins at ${i,j}$ and ${i+1,j}$. They found
that for $0.4\lesssim J_2/J_1\lesssim0.5$, $D_y$ is nearly zero, whereas
$D_x$ only goes to zero at the critical point. These results suggest that
the dimer phase consists of strongly correlated two-chain ladders, which
are then weakly correlated from one ladder to next. This striped nature
of spin correlations in the dimer phase has not been noted before and
is clearly a very interesting result. The situation for $J_2/J_1\gtrsim 0.5$
is again less clear. As discussed before, there are
many possibilities for the phase diagram in that region and much longer series are
needed to throw more light on the situation. Perhaps there is an
interesting multicritical point in that region of the phase diagram.

\section{Discussion}
We have attempted to further elucidate the nature of the intermediate,
magnetically disordered, phase of the spin-$\case 1/2$ $J_1\mbox{-}J_2$ Heisenberg
antiferromagnet on the square lattice. This phase is believed to occur in the
range $0.4 \lesssim  J_2/J_1 \lesssim 0.6$. Our approach has been to derive perturbation
expansions (up to order $\lambda^7$) for the ground state energy, singlet
and triplet excitation energies, and various susceptibilities,
starting from a system of decoupled plaquettes ($\lambda=0$) and extrapolating
to the homogeneous lattice ($\lambda=1$). We have also derived
expansions about an unperturbed state of isolated columnar dimers
(``dimer phase"). Both of these have been proposed as candidates
for the intermediate phase.

We find that the ground state energy for both plaquette and 
dimer phases are very similar, any difference lying within the error 
bars. From this result alone we cannot favor one phase over the other. 

The analysis of the singlet and triplet excitation spectra suggests
an instability in the plaquette phase. 
In particular, in the disconnected
plaquette expansions, Dlog Pad\'e analysis indicates that
the gaps would vanish for $\lambda$ less than unity.
The gap appears to close first for the triplets and then for the singlets.
This is the strongest evidence that the plaquette phase is {\it not}
realized in this model. However, we should mention here that
the critical exponents associated with
the vanishing of the gaps are rather small ($<0.4$) and the gap 
closes not too far from $\lambda$ equal to unity. Thus, with a relatively
short series, this should be treated with some caution.
One could ask why the energy series appear
to converge well despite the instability. However, this is a well known
feature of series expansions, that quantities having weak singularities
may continue to show reasonable values even if extrapolated past the
singularity.

A consistent interpretation of these results is that 
within the parameter space of our non-uniform
Hamiltonian, the plaquette phase is first unstable to a magnetic
phase, which then must give way to the columnar dimer phase. Similar
results for the instability of the staggered dimer phase were suggested
before by Gelfand et al\cite{gel89}. 
However, the full phase diagram in the $J_2/J_1$ and $\lambda$ parameter space
is difficult to obtain reliably, especially near the
transition to the columnar phase. There are possibilities of some
novel multicritical points, which deserve further attention.

One of our most interesting results is the finding of striped spin
correlations in the dimer phase. In this phase, the nearest neighbor
spin correlations are nearly equal along the rungs and along the
chains of a two spin column and there are extended bond correlations
along the chains. However, spin correlations from one column to the next are
much weaker. In other words, the dimers are strongly resonating
along vertical columns. The existence of a quantum critical point
separating an antiferromagnetic phase with such a quantum
disordered phase with striped correlations is a very interesting
feature of this model which deserves further attention in the context
of high-$T_c$ materials.

\acknowledgments
We would like to thank Subir Sachdev and Oleg Sushkov for many
useful discussions.
This work has been supported in part by a grant
from the National Science Foundation (DMR-9616574) (R.R.P.S.), the
Gordon Godfrey Bequest for Theoretical Physics at the University of
New South Wales, and by the Australian Research Council (Z.W., C.J.H. and J.O.).
The computation has been performed on Silicon Graphics Power
Challenge and Convex machines. We thank the New South Wales
Centre for Parallel Computing for facilities and assistance
with the calculations.

% \newpage

%=======================================================================
\begin{figure}[h] %h: here; t:top of page; b:bottom of page; p: page of float
%\vspace{9pt}
\par
\centerline{\hbox{\psfig{figure=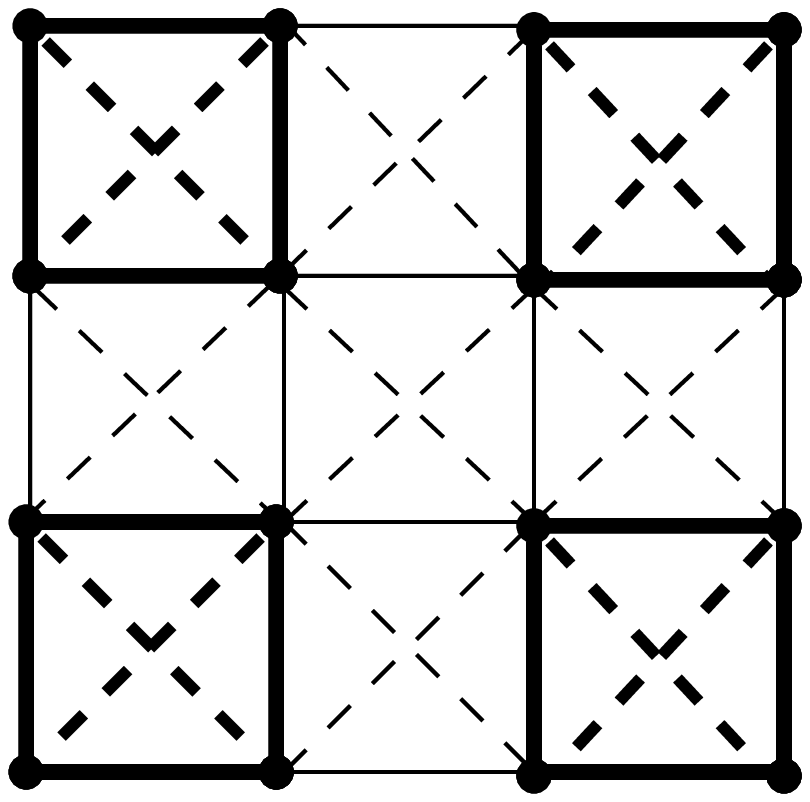,width=5cm}}}
%\par
\vspace{10pt}
\caption{The generalized $J_1$-$J_2$ Heisenberg model with plaquette structure, 
the couplings $J_1$, $J_2$,
$x J_1$, and $x J_2$ indicated  by thick solid
bonds, thick dashed bonds, thin solid bonds and thin dashed
 bonds, respectively.
}
\label{fig_lat}
\end{figure}
%=======================================================================

%=======================================================================
\begin{figure}[h] %h: here; t:top of page; b:bottom of page; p: page of float
%\vspace{9pt}
\par
\centerline{\hbox{\psfig{figure=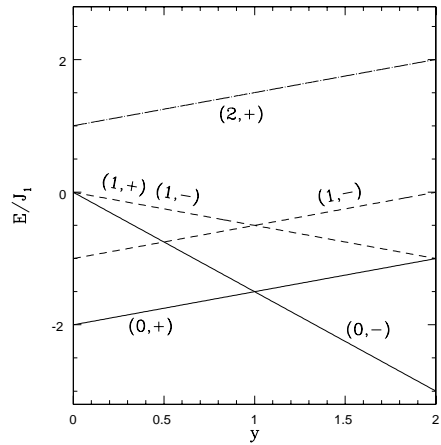,width=8cm}}}
%\par
\vspace{-2pc}
\caption{The energies of an isolated plaquette as function of $y$.
The notation presents the $S$-value and effect of $\pi/2$ rotation $(\pm)$
for each state.
}
\label{fig_eigenvalue}
\end{figure}
%=======================================================================

%=======================================================================
\begin{figure}[h] %h: here; t:top of page; b:bottom of page; p: page of float
%\vspace{9pt}
\par
\centerline{\hbox{\psfig{figure=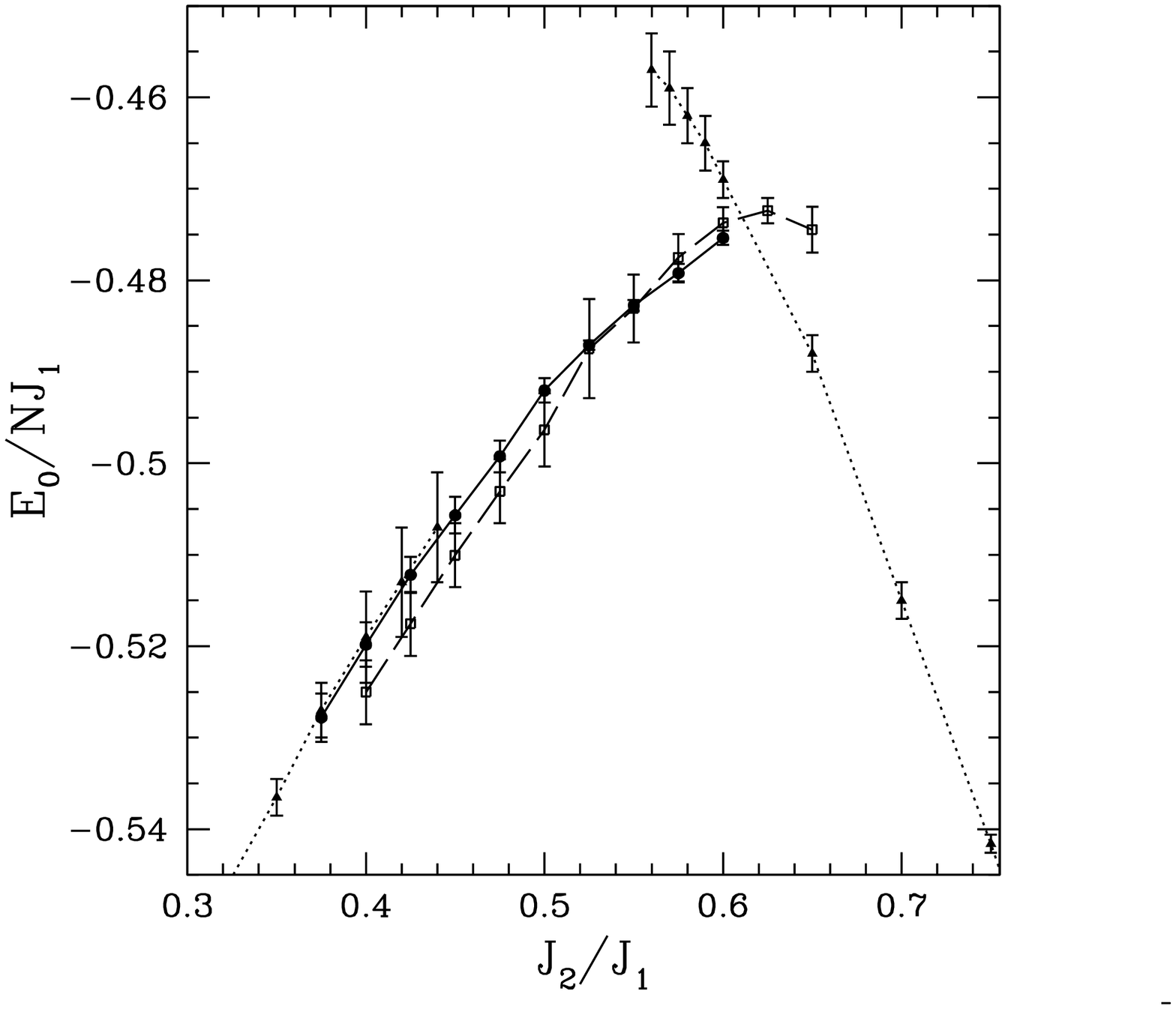,width=8cm}}}
%\par
\vspace{-2pc}
\caption{The ground-state energy per site $E_0/NJ_1$  as
function of $J_2/J_1$, obtained from the Ising expansions\protect\cite{Isingexp}
(triangle points connect by dotted lines), 
the columnar dimer expansions\protect\cite{dimerexp} (open square points connected
by dashed line), 
and the plaquette expansion (full circle points connected by solid line).
}
\label{fig_e0}
\end{figure}
%=======================================================================

%=======================================================================
\begin{figure}[h] %h: here; t:top of page; b:bottom of page; p: page of float
%\vspace{9pt}
\par
\centerline{\hbox{\psfig{figure=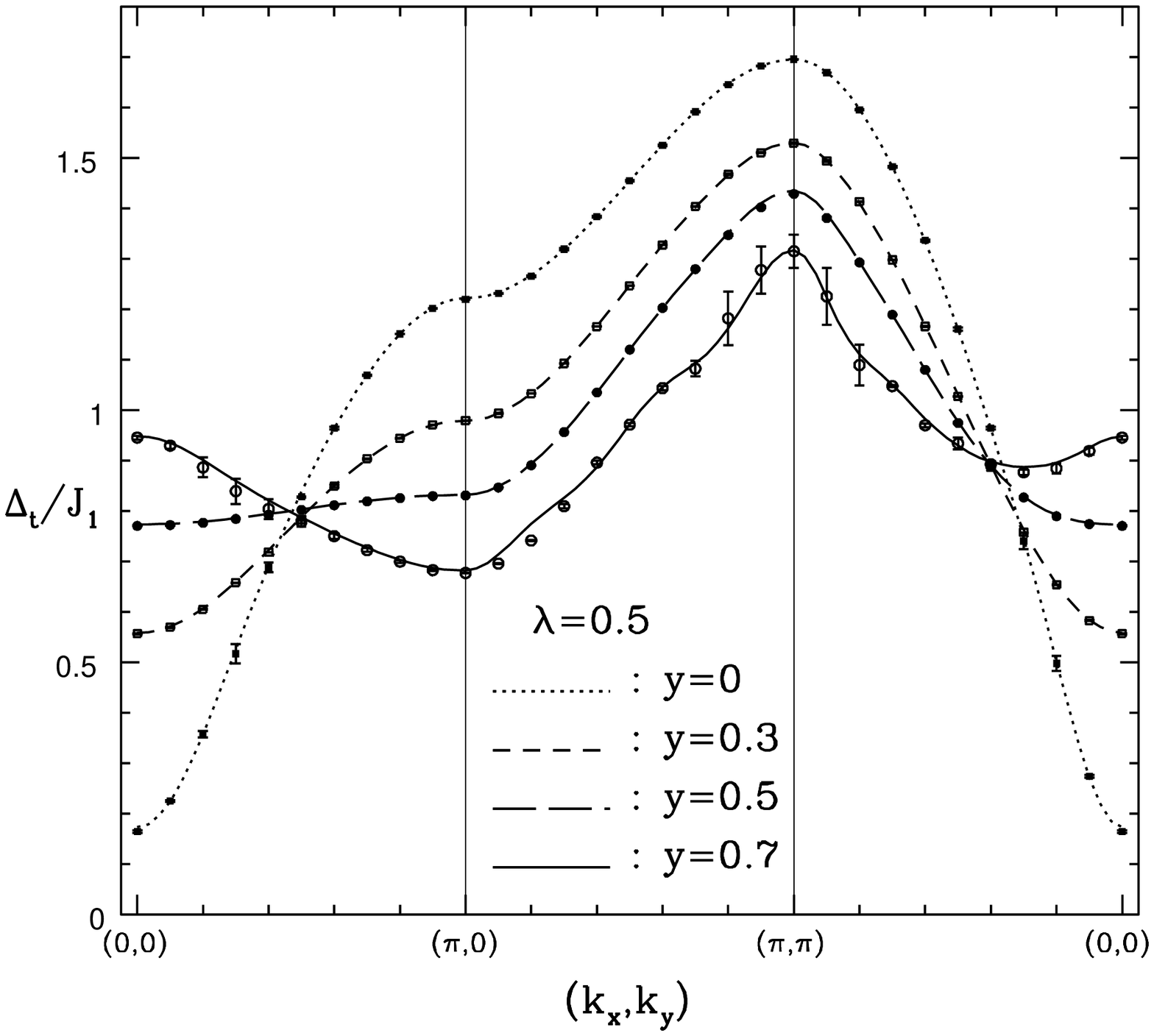,width=8cm}}}
%\par
\vspace{-2pc}
\caption{Plot of the triplet excitation spectrum
$\Delta (k_x , k_y )$ along high-symmetry cuts through the Brillouin
zone for the system with coupling ratios $y=0,0.3,0.5,0.7$ and
$\lambda=0.5$ [shown in the figure from the top to the bottom at
$(\pi,\pi)$, respectively]; the lines are the estimates by direct sum to the
series, and the points  with error bar are the estimates of the integrated
differential approximants to
the series.
}
\label{fig_mk_triplet}
\end{figure}
%=======================================================================

%=======================================================================
\begin{figure}[h] %h: here; t:top of page; b:bottom of page; p: page of float
%\vspace{9pt}
\par
\centerline{\hbox{\psfig{figure=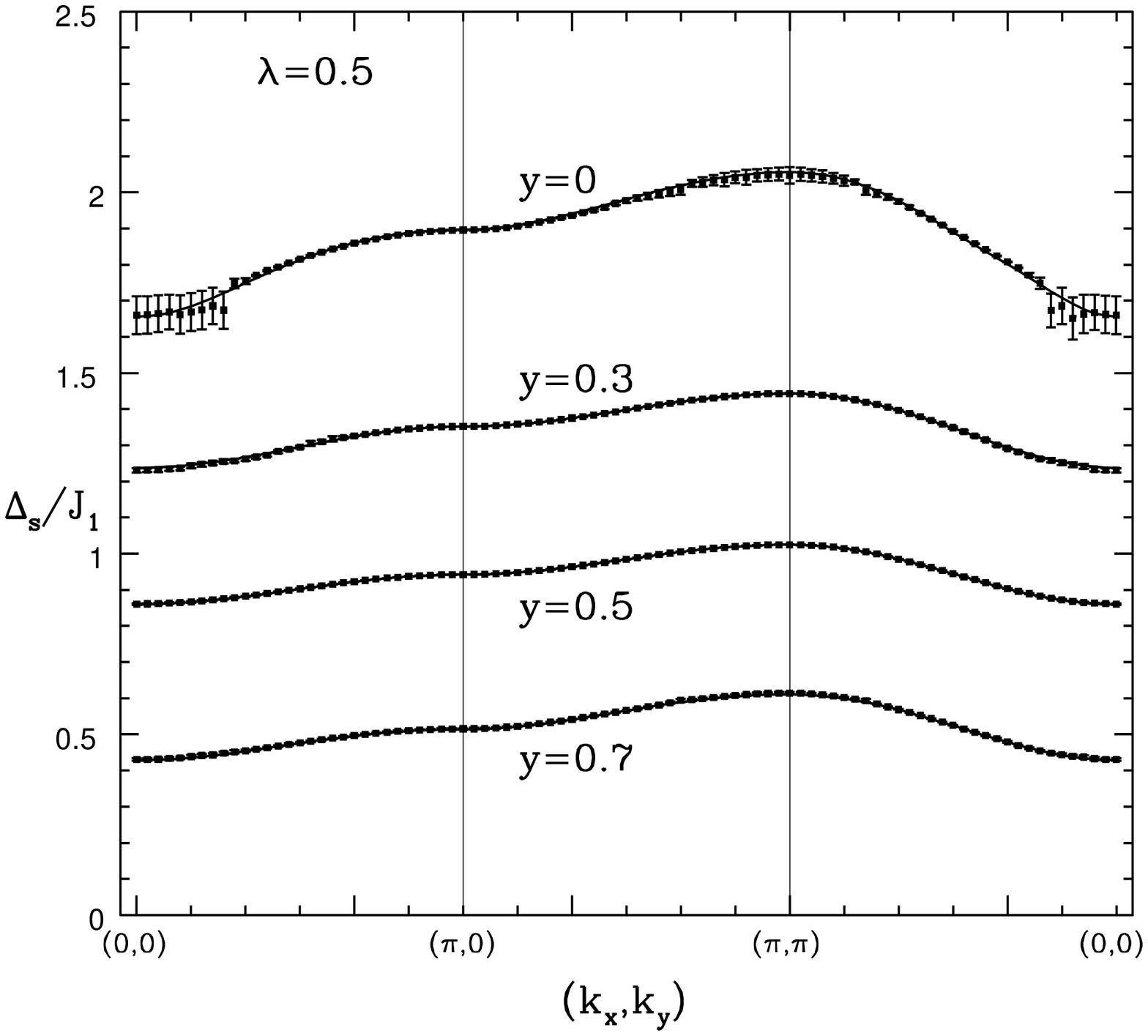,width=8cm}}}
%\par
\vspace{-2pc}
\caption{Plot of the singlet excitation spectrum
$\Delta (k_x , k_y )$ along high-symmetry cuts through the Brillouin
zone for the system with coupling ratios $y=0,0.3,0.5,0.7$ and
$\lambda=0.5$ [shown in the figure from the top to the bottom, respectively]; 
the lines are the estimates by direct sum to the
series, and the points  with error bar are the estimates of the integrated
differential approximants to
the series.
}
\label{fig_mk_singlet}
\end{figure}
%=======================================================================

%=======================================================================
\begin{figure}[h] %h: here; t:top of page; b:bottom of page; p: page of float
%\vspace{9pt}
\par
\centerline{\hbox{\psfig{figure=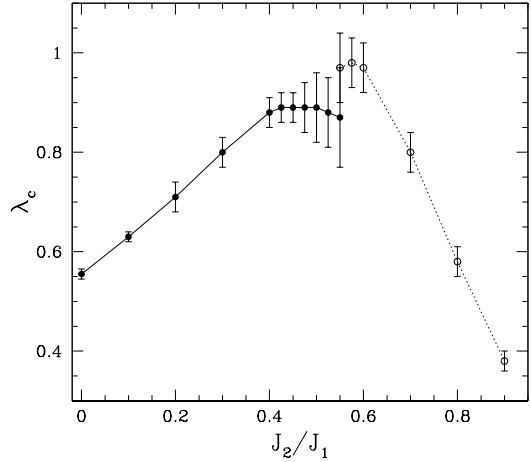,width=8cm}}}
%\par
\vspace{-2pc}
\caption{Phase diagram for generalized $J_1$-$J_2$ Heisenberg model 
with plaquette structure, as determined from the plaquette expansions.
The full (open) points with error bars and a solid (dotted) line to guide
the eye indicate  the line where  the
$(0,0)$ triplet (singlet) gap vanishes.
}
\label{fig_phase}
\end{figure}
%=======================================================================

%=======================================================================
\begin{figure}[h] %h: here; t:top of page; b:bottom of page; p: page of float
%\vspace{9pt}
\par
\centerline{\hbox{\psfig{figure=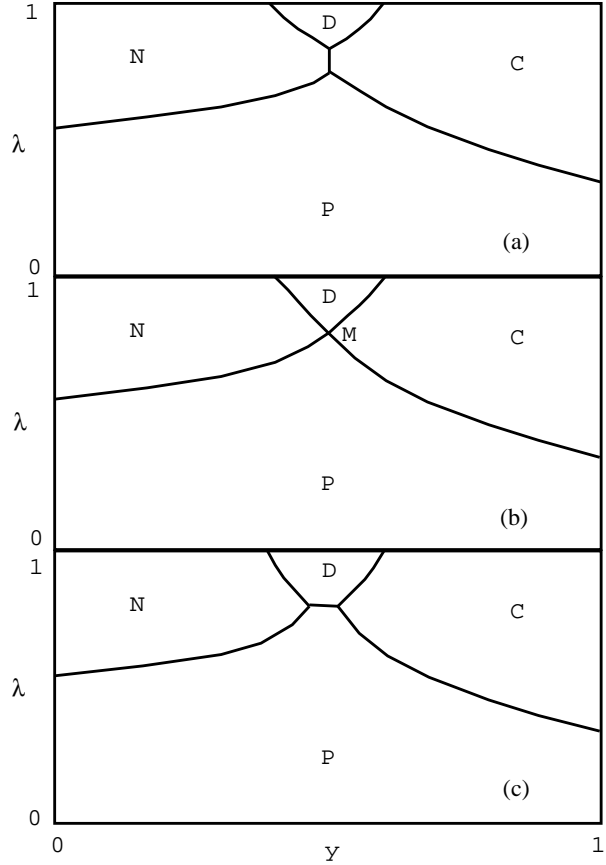,width=8cm}}}
%\par
\vspace{10pt}
\caption{Some possible topologies for the phase diagram in the $\lambda$-$y$ plane. Here, $N$ is
the N\'eel phase, $C$ is the columnar phase, $D$ is the dimer phase and $P$ is the
disordered Plaquette phase with no long range order. Note that some of the phase
boundaries such as those between Plaquette and Dimer and between Dimer and Columnar phases
could be first order, whereas the others could be second order. 
$M$ is a possible multicritical point, where several critical lines meet.
}
\label{fig_pha}
\end{figure}
%=======================================================================

\end{document}